# The AC multi-harmonic magnetic susceptibility measurement setup at the LNF-INFN

Shenghao Wang[1,a], Augusto Marcelli[2,1,b] *, Daniele Di Gioacchino[2,c] and Ziyu Wu[1,d] *

[1]National Synchrotron Radiation Laboratory, University of Science and Technology of China, Hefei, 230026, P.R. China

[2]Istituto Nazionale di Fisica Nucleare - Laboratori Nazionali di Frascati, P.O.Box 13, 00044 Frascati (RM), Italy

[a]ronaldo9@mail.ustc.edu.cn, [b]Augusto.Marcelli@lnf.infn.it, [c]daniele.digioacchino@lnf.infn.it, [d]wuzy@ustc.edu.cn

**Abstract.** The AC magnetic susceptibility is a fundamental method in materials science, which allows to probe the dynamic magnetic response of magnetic materials and superconductors. The LAMPS laboratory at the *Laboratori Nazionali di Frascati* of the INFN hosts an AC multi-harmonic magnetometer that allows performing experiments with an AC magnetic field ranging from 0.1 to 20 Gauss and in the frequency range from 17 to 2070 Hz. A DC magnetic field from 0 to 8 T produced by a superconducting magnet can be applied, while data may be collected in the temperature range 4.2-300 K using a liquid He cryostat under different temperature cycles setups. The first seven AC magnetic multi-harmonic susceptibility components can be measured with a magnetic sensitivity of $1 \times 10^{-6}$ emu and a temperature precision of 0.01 K. Here we will describe in detail about schematic of the magnetometer, special attention will be dedicated to the instruments control, data acquisition framework and the user-friendly LabVIEW-based software platform.

**Introduction**

The AC magnetic susceptibility measurement is a unique method suitable to provide a precise characterization of magnetic and superconducting materials in a non-destructive way, particularly appropriate to study the dynamic magnetic response of a material [1]. In an AC magnetic susceptibility measurement experiment, the magnetization is periodically changed in response to a time-varying exciting magnetic field $h(t)$:

$$h(t) = h_0 \cos(\omega t). \tag{1}$$

whose magnetization oscillations can be described as:

$$M(t) = \chi' h_0 \cos(\omega t) + \chi'' h_0 \sin(\omega t). \tag{2}$$

where χ' and χ" represent respectively the in-phase and out-of-phase components of the magnetic susceptibility with the applied magnetic field. The in-phase component χ' is associated to the dispersive magnetic response of the sample, while the imaginary part χ" is proportional to the energy dissipation. The latter is converted into heat during one cycle of the AC field or the energy absorbed by the material from the applied AC field. Using a complex notation, χ' and χ" can be combined to form the complex susceptibility. If the system has a magnetic linear response, only the fundamental sinusoidal exciting frequency waveform is present. However, when the magnetic response of the sample is non linear (e.g., for magnetic or superconductor samples), the pure sinusoidal field induces a distorted waveform characterized by non-sinusoidal oscillations of the magnetization of the material. These oscillations may be described as a sum of sinusoidal components that oscillate at harmonics of the driving frequency:

$$M(t) = h_0 \sum [\chi'_n \cos(n\omega t) + \chi''_n \sin(n\omega t)] . \qquad (3)$$

where $\chi'_n$ and $\chi''_n$ (n = 1, 2, 3…) are the in-phase and out-of-phase components of the harmonic susceptibilities. In the complex notation, $\chi'_n$ and $\chi''_n$ can be combined to form the complex harmonic susceptibilities.

$$\chi_n = \chi'_n + i\chi''_n . \qquad (4)$$

A reliable characterization of materials with a non-linear magnetic response requires the measurement of harmonic susceptibilities beyond the fundamental one. Moreover, the AC magnetic multi-harmonic susceptibility, by probing the real and imaginary parts of the first and the higher harmonic susceptibility, may separate linear and non-linear transport processes occurring in a material, allowing the determination of different magnetic/superconducting phases eventually present in the sample under analysis.

A multi-harmonic AC magnetic susceptibility measurement setup is operational in the LAMPS laboratory (*LAboratory for Magnetism High Pressure and Spectroscopy*) of the Laboratori Nazionali di Frascati (LNF) of the Istituto Nazionale di Fisica Nucleare (INFN). It was built from commercial available components and custom made devices, these scientific instruments come from different software background. To measure the susceptibility, users need to change the temperature thousands of times during an acquisition and it is almost impossible to record data and display them using the original instrumental interfaces, controls and data acquisition packages. In order to optimize the management of instruments, and to realize automatic data acquisition, display and recording, a dedicated software platform was supposed to get developed.

As a graphical programming language, LabVIEW (National Instrument's Laboratory Virtual Instrumentation Engineering Workbench) is now a very popular development tool widely used in many scientific and industrial areas. The easy availability of hardware drivers for a very large number of scientific instruments, the easy-to-use multithreaded programming, the convenient graphic user interface (GUI) design, the high-efficiency debugging functions and many other remarkable features make LabVIEW a flexible tool optimized for an instrument-oriented programming environment. Actually, it has been already successfully used in many other similar applications [2-4].

Here we will describe schematically the diagram and the working principle of the AC magnetic susceptibility measurement setup available at LAMPS. The control of the instruments, the data acquisition framework and the user-friendly LabVIEW-based software platform will also be introduced. Finally, as an example, experimental results performed on a superconducting pnictide sample using this measurement setup will be presented.

**The measurement setup and the working principle**

In the AC magnetic susceptibility setup, the magnetic moment and the susceptibility of the sample are measured by the induction in two counter-wound coils while the AC-magnetic field is applied. Indeed, the voltage induction in the coils is directly related to the magnetic moment in the sample. Fig. 1 illustrates the layout of the AC magnetometer. The heart of the instrument is a gradiometer based on a bridge made by two pick-up coils connected in series and wounded in the opposite sense. It is surrounded by a driving excitation coil, which is called the first derivative configuration of the gradiometer coil. The design is used to reduce the magnetic field fluctuations noise in the detection circuit due to the applied magnetic field. The excitation coil receives the AC signal from a function generator, magnified by an amplifier that yields to an alternating driving magnetic field, whose frequency may range from 17 Hz to 1070 Hz with a variable amplitude from 0 to 20 Gauss.

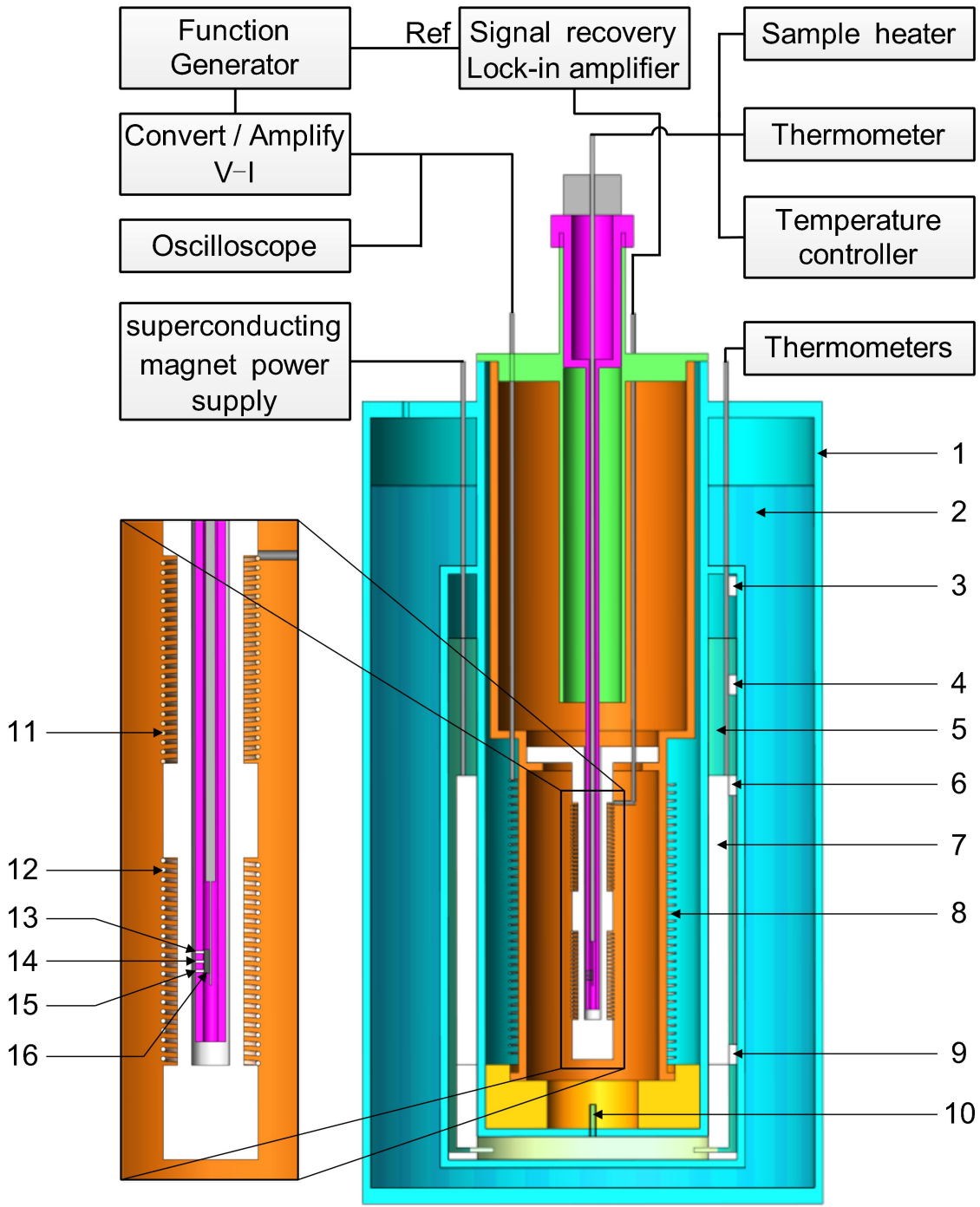

Figure 1. Layout of the AC magnetic susceptibility setup. 1.double vessel cryostat, 2. liquid $N_2$, 3. carbon resistor inside the chamber, 4. carbon resistor at the VTI position, 5. liquid He, 6. carbon resistor on top of the magnet, 7. magnet, 8. excitation coil, 9. carbon resistor at the bottom of the magnet, 10. pinhole, 11. pick-up bridge (balance) coil, 12. pick-up bridge (sensing) coil, 13. Pt thermometer, 14. carbon resistor thermometer, 15. carbon resistor (sample heater), 16. sample.

The sample is mounted on a sapphire holder slab located at the centre of one of the two pick-up coils (the "sensing" coil) of the bridge while the second one (the "balance" coil) has to remain empty. Being non-magnetic and a good thermal conductor, sapphire was chosen as the "substrate". Also importantly, characterized by a low electric conductivity, sapphire guarantees low current losses. A Pt thermometer and a carbon resistor placed near the sample are in thermal contact with the sapphire holder. The first reads the temperature of the sample while the carbon resistor is used to heat the sample in order to perform experiments vs. temperature in a range going from 4.2 to 300 K.

The AC magnetic field generated by the driving coil induces a variable magnetic moment on the sample and consequently a flux variation in the pick-up coils, whose voltage signal is measured by a multi-harmonic lock-in amplifier. The latter acts as a discriminating voltmeter, it measures the amplitude and the relative phase of the AC signal, using the same fundamental frequency of the excitation as the reference signal while a fixed phase relationship is provided respect to it. The lock-in amplifier is a band pass filter with a very large Q with its center frequency set at the selected signal frequency. The output is an amplified DC voltage proportional to the synchronous AC input signal. The temperature of the sample holder and the coils assembly can be varied. They are located in a double vessel cryostat thermally controlled by a He gas-flow. The outer vessel is filled with liquid nitrogen, while the liquid He fills the inner one, where a superconducting magnet operates in the range from 0 to 8 T in persistent or non-persistent modes. The sample is mounted at the centre of the superconducting magnet and its temperature can be changed via the cold He gas flow from a pinhole that is manually controlled via a throttle and a needle valve on the top of the cryostat. Before any measurement a purging of the sample compartment with clean He gas has to be done.

The magnetic measurements can be performed both in the zero field cooled (ZFC) and in the field cooled (FC) modes. In the ZFC mode the sample is slowly cooled below the transition temperature without the DC magnetic field, then the magnetic field is turned on, while in the FC mode the magnetic field is turned on above the critical temperature $T_c$ of the superconducting or of the magnetic phase. Then the sample is cooled down below the transition temperature and after that, the temperature is let to increase and the measurement starts.

AC magnetic multi-harmonic susceptibility experiments with a magnetic sensitivity down to $1 \times 10^{-6}$ emu and a temperature precision of 0.01 K can be performed with this setup. Higher components of the magnetic susceptibility can be collected up to the seventh harmonic. All these characteristics make this instrument particularly suitable to probe weak magnetic phases present in diluted magnetic structures such as antiferromagnet, super paramagnetic systems, spin glass and the transport properties of many complex materials and in particular, to investigate the vortex dynamics of superconductor materials.

**Instrumental control and data acquisition**

Fig. 2 shows the instrumental control and the data acquisition scheme of the magnetic susceptibility measurement setup. We utilize HP 8116A function generator to produce a sinusoidal signal, which is magnified by a custom made amplifier. The released sinusoidal signal is monitored by the Tektronix TDS 1002 oscilloscope, and then the signal is transmitted to the excitation coil.

The signal recovery 7265 multi-harmonic lock-in amplifier is used here to measure the voltage signal generated by the flux variation in the pick-up coils. The temperature of the sample under analysis is collected by the Lake Shore 218 thermometer and the Oxford temperature controller ITC503. The DC superconducting magnet is controlled by the Oxford power supply IPS120-10. An Agilent 34970A data acquisition switch unit is dedicated to measuring the resistance of the four carbon resistors mounted at four reference positions of the liquid He tank, by which we could evaluate continuously the level of the liquid He inside the reservoir.

Communications of all instruments and their management with a personal computer were realized in different ways following correlated protocols schematically illustrated in Fig. 2. This multipurpose configuration includes two personal computers, one located near the instruments and one in the control room, both should be capable to run data acquisition and to control instruments. The mission is fulfilled by means of special network communication devices and the help of a serial server MoxaNport 5410. Moreover, the original RS-232 serial communication mode of the superconducting magnet power supply was transferred to the Ethernet-based (LAN) TCP/IP protocol that makes this instrument also available on the network. A high-performance Ethernet-to-GPIB controller (NI GPIB-ENET/1000), with a maximum GPIB transfer rate of 5.6 MB/s supports the access via TCP/IP to all aforementioned instruments. Finally, the Asus GX1008 is the switchboard that allows to share all instruments among two users on the network by hinging together the serial server, the Ethernet-to-GPIB controller and up to two personal computers.

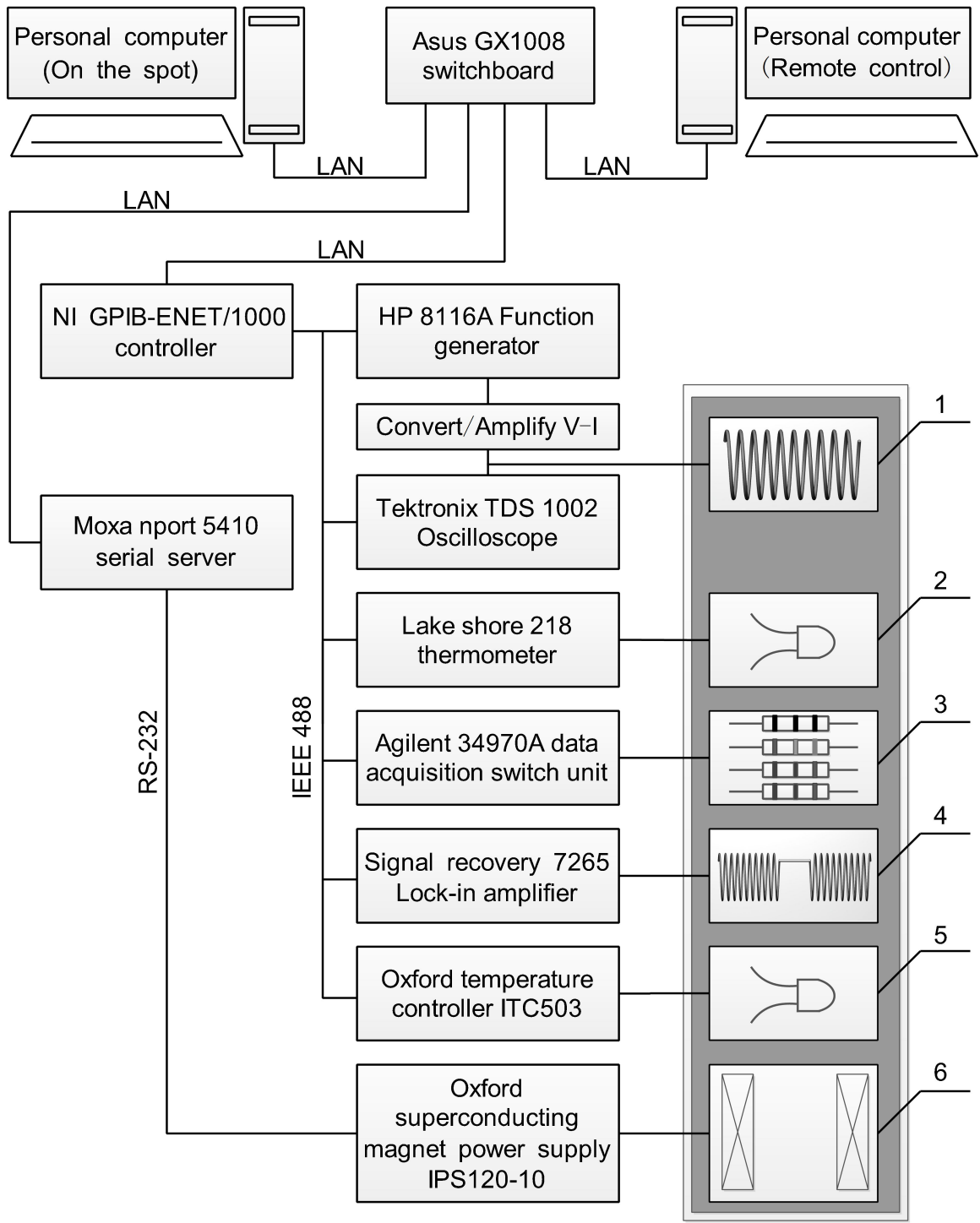

Figure 2. Control and data acquisition logic of the magnetic susceptibility setup. 1. excitation coil, 2. Pt thermometer, 3. carbon resistors, 4. pick-up bridge coil, 5. carbon resistor, 6. magnet.

**The LabVIEW-based software platform**

Fig. 3 is the main GUI of the LabVIEW based software platform. In the next we will describe how a typical experiment run. After starting the program, users need to login from the main menu, help and contacts information are also accessible from the same menu. In the second phase users would be asked whether to enable the superconducting magnet field. The positive answer starts the module of superconducting magnet controlling, while the program jumps to the next procedure if you likely run experiments with the DC magnet field sets to zero. In the following step, users are required to set experimental parameters and instruments: file name of the recorded data, measurement time interval and a valid temperature range for the sample. On the right part of the GUI we found how to set each instrument. Experimental notes can be written down by users in the bottom of the GUI.

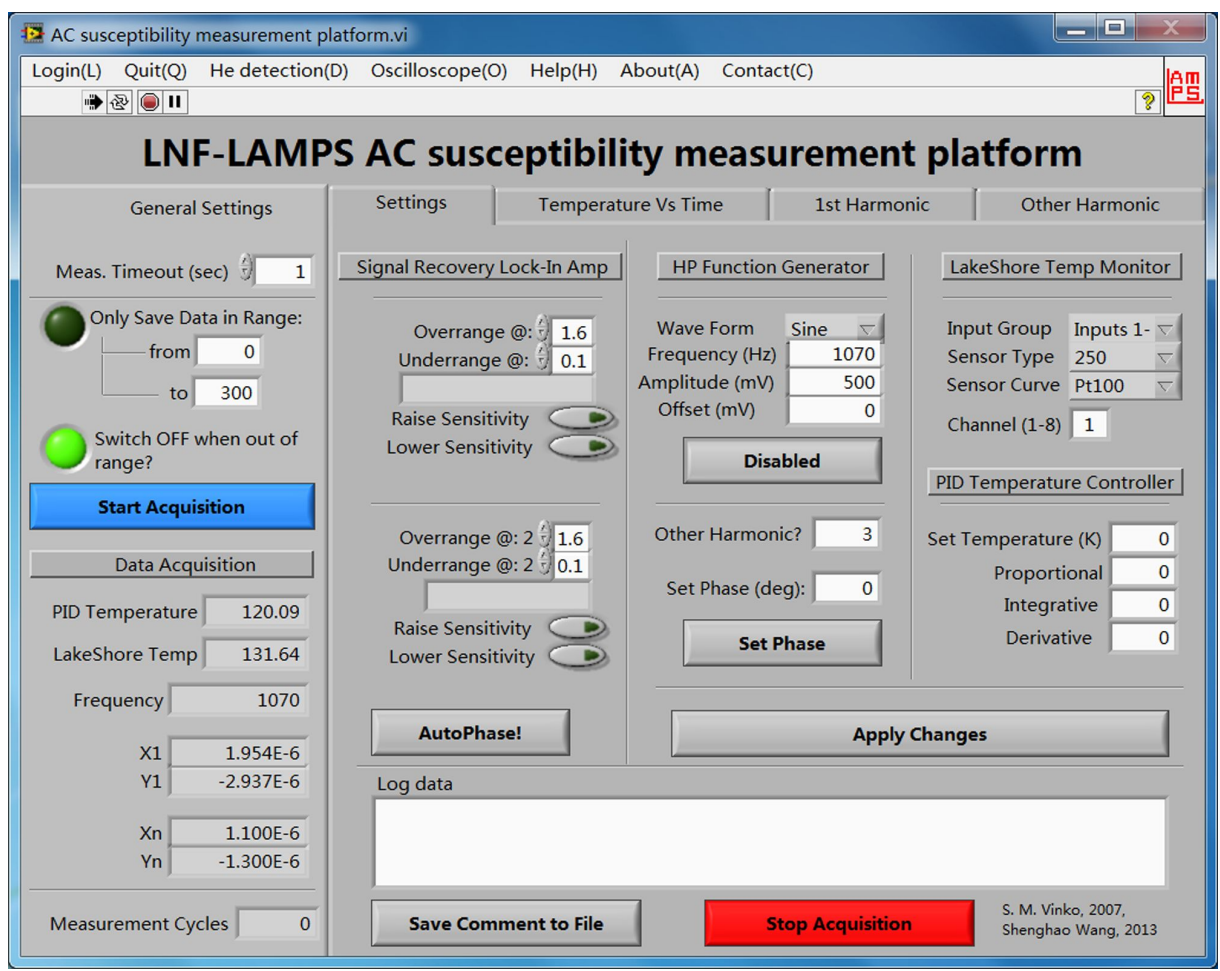

Figure 3. Main GUI of the AC susceptibility measurement software platform.

After correct setting of the required parameters, the "Start Acquisition" button triggers initialization of all the involved devices. Then, data would be acquired from the temperature controller, thermometer and signal recovery, at the same time, graph of temperature vs. time, together with the 1st harmonic and the other harmonics vs. temperature would be displayed in different pages of the Tab control. And measurement results will be stored (with a featured format) as a text file that allow users to conduct off-line data post-processing. From main GUI of the software platform, measurement of the level of the liquid He and monitoring of the excitation signal waveform are available via the "He detection" and "Oscilloscope" menu, respectively. More details about the software development procedure, the flow chart and other features can be found in Ref. [5].

**Experimental results and discussion**

To give a better feeling of the instrument, we will show AC multi-harmonic susceptibility data collected on the $N_dFeAsO_{1-0.14}F_{0.14}$ compound using the above experimental setup and software. The sample was synthesized in Beijing by a high-pressure synthesis method starting from Nd, As, Fe, $Fe_2O_3$, $FeF_3$ powders. Additional information of this sample are available in Ref. [6]. In the experiments, the sample was cooled with a zero field cooling (ZFC) procedure, the amplitude of the applied AC magnetic field was set to 3.9 G and data were collected at the frequency of 507 Hz.

First and third harmonics of the AC magnetic susceptibility were recorded and datas are shown in Fig. 4. From the behaviour of the first and third harmonics of the AC magnetic susceptibility vs. temperature, we recognized a superconducting diamagnetic phase around $T_c$ (47 K) and an antiferromagnetic phase around $T_m$ (90 K). Spectra are clearly separated, in particular in the plot of the real part of the first harmonic $\chi'_1$. Moreover, the imaginary part of the first harmonic $\chi''_1$ shows an increase of the area around the phase transitions. Both the real and the imaginary parts of the third harmonic, named $\chi'_3$ and $\chi''_3$, respectively, confirm that both phase transitions occur. The measurement points out also that in the $N_dFeAsO_{1-0.14}F_{0.14}$ compound the anti-ferromagnetic and the superconducting phase coexist, a phenomenon unusual in a superconductor material. This observation could be explained here by the presence of an exchange field of local Nd *4f* spins (AF phase) and *3d* Fe electron pairs in the FeAs layer (superconducting phase) of this pnictide [7].

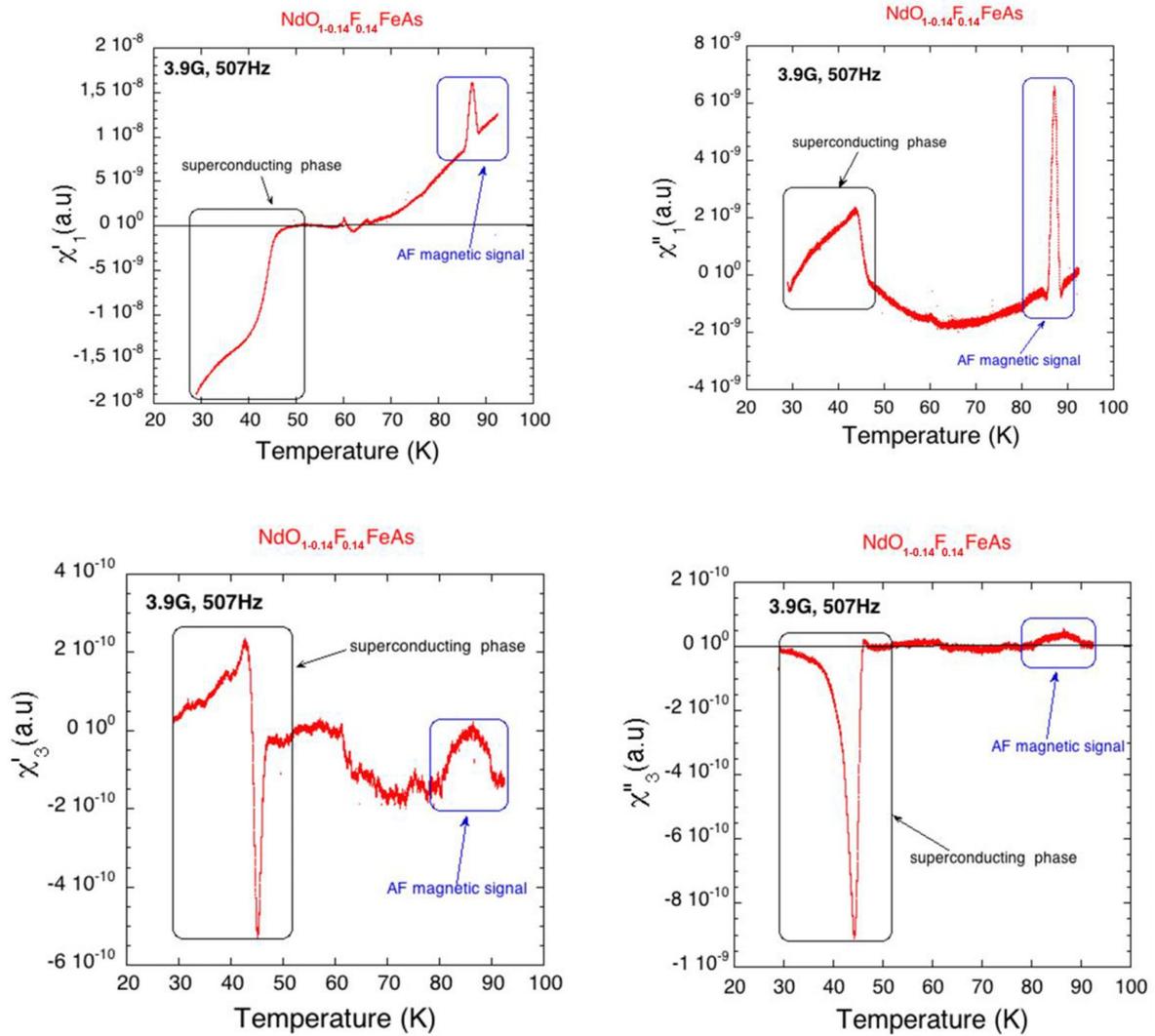

Figure 4. AC multi-harmonic susceptibility of $N_dFeAsO_{1-0.14}F_{0.14}$. (top): real ($\chi'_1$) and imaginary part ($\chi''_1$) of the first harmonic; (bottom): real ($\chi'_3$) and imaginary part ($\chi''_3$) of the third harmonic.

**Conclusion**

In this manuscript we described the AC magnetic susceptibility measurement setup available at the LAMPS laboratory of INFN-LNF, this system is running at Frascati since 1998 and many fruitful scientific achievements have been obtained [8-14]. In this contribution a special attention has been devoted to describe its instruments control, data acquisition framework and the LabVIEW-based user-friendly, convenient software platform.

**Acknowledgements**

The authors would greatly thank Dr. Xing Chen for fruitful discussion and support. This work was partly supported by the National Basic Research Program of China (2012CB825800), the Science Fund for Creative Research Groups (11321503), the Knowledge Innovation Program of the Chinese Academy of Sciences (KJCX2-YW-N42), the National Natural Science Foundation of China (NSFC 11179004, 10979055, 11205189, and 11205157) and the Fundamental Research Funds for the Central Universities (WK2310000021).